\documentclass{aa}     
\usepackage{graphicx}
\usepackage{txfonts}
\usepackage{natbib}
\bibpunct{(}{)}{;}{a}{}{,} 
\newcommand{\Mpc}{$h^{-1}$\,Mpc}

\begin{document}

\title{Anatomy of luminosity  functions: the 2dFGRS example}

\author {E.~Tempel \and J.~Einasto \and
  M.~Einasto \and E.~Saar \and E.~Tago
}

\institute{Tartu Observatory, EE-61602 T\~oravere, Estonia
}
\date{ Received 28 May  2008 / Accepted  2008  }

\authorrunning{E.~Tempel et~al.}
\titlerunning{2dF luminosity function}

\offprints{E.~Tempel (\email{elmo@aai.ee}) }

\abstract
{}
{ We use the 2dF Galaxy Redshift Survey to derive the luminosity
  function (LF) of the first-ranked (brightest) group/cluster galaxies, the LF
of
  second-ranked, satellite and isolated galaxies, and the LF of groups
  of galaxies.}
{ We investigate the LFs of different samples in various environments:
  in voids, filaments, superclusters and supercluster cores. We
  compare the derived LFs with the Schechter and double-power-law
  analytical expressions. We also analyze the luminosities of isolated
  galaxies.}
{ We find strong environmental dependency of luminosity functions of
  all populations. The luminosities of first-ranked galaxies have a
  lower limit, depending on the global environment (higher in
  supercluster cores, and absent in voids). The LF of second-ranked
  galaxies in high-density regions is similar to the LF of first-ranked galaxies
in a lower-density environment.  The brightest
  isolated galaxies can be identified with first-ranked galaxies at
  distances where the remaining galaxies lie outside the observational
  window used in the survey.}
{ The galaxy and cluster LFs can be well approximated by a double-power-law;
  the widely used Schechter function does not describe well the bright
  end  and the bend of
  the LFs. Properties of the LFs reflect differences in the evolution of 
  galaxies and their groups in different environments.}


\maketitle

\section{Introduction}

Groups and clusters of galaxies are the most common environment of galaxies.
In particular, groups of galaxies are locations of galaxy formation, and their
study yields information on the processes of galaxy formation and evolution.
Clusters of galaxies form basically by hierarchical merging of smaller units
-- galaxies and groups of galaxies.  In groups and clusters the evolution of
galaxies differs from that in low-density regions.

The presence of satellite galaxies around our Galaxy and the Andromeda
galaxy is known long ago.  Systematic studies of physical groups of
galaxies were pioneered by
\citet{holmberg:69,devaucouleurs:70,turner:74}, followed by
\citet{geller:83,nolthenius:87,tully:87,maia:89,ramella:89,gourgoulhon:92,
  garcia:93,moore:93}, and many others.  First large catalogues of
clusters of galaxies were constructed by visual inspection of the
Palomar Observatory Sky Survey plates by
\citet{abell:58,zwicky:68,abell:89}.  More recently clusters of
galaxies were selected also using their X-ray emission by
\citet{gioia:90,ebeling:96,bohringer:01}.  Deeper redshift surveys
made it possible to construct group and cluster catalogues for more
distant objects: e.g., the Las Campanas Redshift Survey was used by
\citet{tucker:00} to construct a catalogue of loose groups.  The
'100k' public release of the 2 degree Field Galaxy Redshift Survey
(2dFGRS), described by \citet{colless:03}, has been used to construct
several catalogues of groups.  Among them, there are the catalogues by
\citet{merchan:02}, by \citet{yang:05a}, and by \citet[][hereafter
T06]{tago:06}.  \citet{eke:04a} used the complete 2dFGRS to compile a
sample of about 25 thousand groups and clusters in the two contiguous
Northern and Southern Galactic Patches.

One of the principal description functions for clusters and groups of
galaxies is the luminosity function $F(L)$ that describes the average
number of galaxies per unit volume as a function of galaxy
luminosity. The luminosity function (LF) plays an important role in
our understanding how galaxies form and evolve
 \citep{vandenbosch:03,yang:03,cooray:05b,cooray:05a,
  cooray:06,milosavljevic:06,tinker:05,tinker:07,hansen:07,vandenbosch:07,
  tinker:08}.

The LF of groups of galaxies was first derived by \citet{holmberg:69},
followed by \citet{christensen:75,kiang:76,abell:77,mottmann:77}.
These studies showed that the LF of galaxies in groups and clusters
can be approximated by a double-power-law, the crossover between two
powers occurs at a characteristic absolute magnitude $M^{*} \approx
-20 - 5\log h$.

Our interest in the structure of groups of galaxies began with the
discovery of dark matter coronas (haloes) around giant galaxies
\citep{einasto:74a}.  We noticed that practically all giant galaxies
are surrounded by dwarf companion galaxies, and that such systems have
a certain structure: elliptical companions are concentrated near the
dominating (brightest) galaxy, and spiral and irregular companions lie
at the periphery of the system \citep{einasto:74b}.  The LF of these
systems has a specific feature: the luminosity of the brightest galaxy
of the system exceeds the luminosity of all companion galaxies by a
large factor, thus the overall relative LF of the system -- the
  conditional luminosity function -- has a gap separating the
brightest galaxy from companion galaxies \citep{einasto:74e}.
Dynamically these systems are dominated by dark matter, and there
exist clear signatures of mutual interactions between galaxies and
intergalactic matter in these systems, as shown by
\citet{einasto:74b,einasto:75,chernin:76,einasto:76a} (for a
  review of dark matter around galaxies see \citet{faber:79}).  The
  morphology-density relation in clusters was investigated by \citet{oemler:74,
    dressler:80, postman:84}.  We see that it is valid also in
  ordinary groups.  In other words, groups of galaxies
are not just random collections of galaxies, they form systems with
various mutual interactions.  The whole system of companion galaxies
lies inside the dark corona (halo) of the brightest galaxy and can be
considered as one physical entity.  To stress this aspect we called
such systems hypergalaxies \citep{einasto:74e}.  Our first catalogue
of hypergalaxies (groups with a dominating brightest galaxy) was
composed by \citet{einasto:77}. In recent years the study of the
  connection between dark matter haloes and galaxies has made great
  progress, in particular using the halo occupation distribution
  model; for details see, among others, \citet{kauffmann:97,tinker:05, yang:05a,
   zehavi:05, tinker:06, zheng:07, yang:08b}.

The dominating role of the brightest (first-ranked) cluster/group
galaxies was known long ago, for early studies see
\citet{hubble:31,hubble:36,sandage:76}.  The nature of physical
processes which influence the luminosity and morphology of galaxies in
clusters (and groups) is also known: tidal-stripping of gas during
close encounters and mergers \citep{spitzer:51}, ram-pressure sweeping
of gas due to galaxy motion through the intra-cluster medium
\citep{gunn:72,chernin:76,vandenbosch:08}, galaxy mergers
\citep{toomre:72}.

To understand details of these processes it is important to study
properties of galaxies in groups and clusters.  Indeed, in last years
the number of studies devoted to the study of LFs in groups and
clusters has increased.  We note here the work by
\citet{ferguson:88,ferguson:91,vandenbergh:92,moore:93,sulentic:94,ribeiro:94,
  zepf:97,
  hunsberger:98,muriel:98,zabludoff:00,popesso:05,miles:04,miles:06,
  gonzalez:05, gonzalez:06, berlind:06,chiboucas:06,lin:06,
  zandivarez:06,adami:07,hansen:07,milne:07,vale:06,vale:08}.

The present analysis has three goals: to determine the LF of group
brightest (first-ranked), second-ranked and satellite galaxies; to
investigate the
nature of satellite and isolated galaxies; and to analyze
environmental dependency of galaxy luminosities.  As there are no
strict differences between groups and clusters of galaxies we shall
use the term ``group'' for groups of galaxies as well as for
  conventional clusters. To derive the LFs we shall use the
catalogue of groups and clusters by Tago et~al. (T06).  This catalogue
was prepared using the 2dFGRS Northern and Southern Galactic Patches,
similar catalogues have been compiled by \citet{eke:04a,yang:05a}
  and several other authors. As in all such
catalogues, we find a number of isolated galaxies, i.e. galaxies which
have no neighbours within the search radius in the flux-limited galaxy
survey.  We analyse the luminosity distribution of isolated galaxies
and show that a large fraction of these galaxies can be considered as
brightest galaxies of groups where fainter members of the group
lie outside the visibility window of the survey. As a by-product we
derive also the LF of groups.

LFs of simulated groups and groups found for the 2dFGRS and the
  Sloan Digital Sky Survey Data Release 4 have been recently studied,
  among others, by \citet{mo:04,yang:04,cooray:05b, cooray:05a,
    croton:05, zheng:05, tinker:08, yang:08a}. In many of these papers
  the emphasis has been on explanation of the LF using halo
  occupation statistics.  Our motivation in this paper is mostly
  observational; we shall study the connection of the observed LF with
  sub-halo model data in a separate paper (in preparation). Here we
  shall discuss the nature of second-ranked and satellite galaxies in
  more detail, and shall search for the dependence of the LFs of the
  brightest and satellite galaxies on the environment.

The observational data are discussed in the next Section; here we
consider also the selection effects and the methods to correct data
for selection.  In Sect.~\ref{sec:3} we calculate the LFs of group
brightest, second-ranked, satellite and isolated galaxies. We derive
the LFs for different environmental densities. The nature of isolated
galaxies is discussed in Sect.~\ref{sec:4}.  The LFs of various galaxy
samples and the LF of groups are derived in Sect.~\ref{sec:5}: here we
compare the Schechter and double-power-law expressions. We discuss our
results and bring conclusions in Sects.~\ref{sec:6} and~\ref{sec:7},
respectively.

\begin{table}
\caption{Data on the 2dFGRS galaxies and groups used}
\label{Tab1}
\begin{tabular}{cccccc}
\hline\hline
\noalign{\smallskip}
  Sample & $\Delta$RA & $\Delta$DEC &
  $N_{\rm gal}$&$N_{\rm gr}$&$N_{\rm isol}$\\ 
 & deg & deg & & & \\
\hline
1 & 2 & 3 & 4 & 5 & 6 \\
\hline
\noalign{\smallskip}
  NGP& $75^{\circ}$ &$10^{\circ}$ & 78067   & 10750   & 44134\\
  SGP& $90^{\circ}$&$13.5^{\circ}$ & 106328  & 14465   & 61344\\
\hline
\end{tabular}

\medskip
Columns:
\begin{list}{}{}
\item[1:] the subsample of the 2dFGRS catalogue.
\item[2:] sample width in right ascension (degrees).
\item[3:] sample width in declination (degrees).
\item[4:] total number of galaxies.
\item[5:] number of groups.
\item[6:] number of isolated galaxies.
\end{list}
\end{table}

\begin{figure}
\resizebox{\hsize}{!}
{\includegraphics{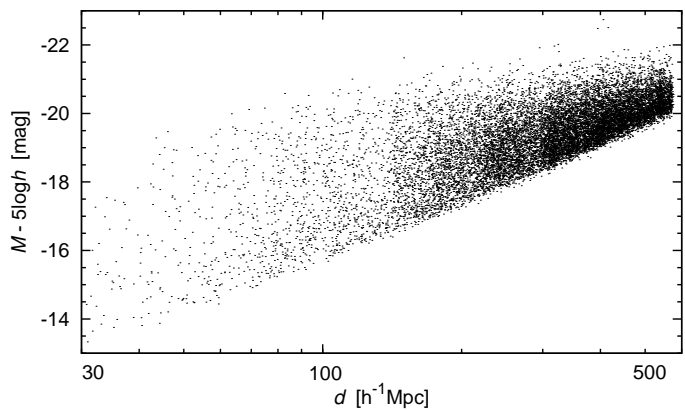}}
\caption{Luminosities of first-ranked galaxies in the 2dFGRS at
  various distances from the observer.  }
\label{fig:1}
\end{figure}

\section{Data}

\subsection{The group catalogue}

In the present analysis we shall use the catalogue of groups and
clusters by Tago et~al. (T06).  This catalogue covers the contiguous
2dFGRS Northern and Southern Galactic Patches (NGP and SGP,
respectively), small fields spread over the southern Galactic cap are
excluded. We extracted data on galaxies from the 2dFGRS web-site
(\texttt{http://www.mso.anu.edu.au/2dFGRS}): the coordinates RA and
DEC, the apparent magnitudes in the photometric system $b_J$, the
redshifts $z$, and the spectral energy distribution parameters $\eta$.
We excluded distant galaxies with redshifts $z > 0.2$, since weights
to calculate expected total luminosities (see Sect.~\ref{sec:totlum})
become too large and uncertain at these redshifts.  The apparent
magnitude interval of the 2dFGRS ranges from $m_1 = 14.0$ to the
survey faint limit $m_2\approx 19.45$ (in the photometric system
$b_J$, corrected for the Galactic extinction).  Actually the faint
limit $m_2$ varies from field to field.  In calculation of the
luminosity weights these deviations have been taken into account, as
well as the incompleteness of the survey (the fraction of observed
galaxies among all galaxies up to the fixed magnitude limit; for
details see T06). The number of galaxies selected for the analysis is
given in Table~\ref{Tab1}.  For linear dimensions we use co-moving
distances \citep[see, e.g.][]{martinez:03}, computed using a
$\Lambda$CDM cosmological model with the following parameters: the
matter density $\Omega_\mathrm{m} = 0.3$, and the dark energy density
$\Omega_{\Lambda} = 0.7$.

In the group definition T06 tried to avoid the inclusion of large
sections of underlying filaments or surrounding regions of
superclusters into groups.  To find the appropriate search radius
(FoF radius) for group definition T06 investigated the behaviour of
roups, if artificially shifted to larger distances from the observer.
Using this method T06 found that the search radius to find group
members must increase with distance only moderately.

To transform the apparent magnitude $m_J$ into the absolute magnitude $M$ we 
use the usual formula
\begin{equation}
 M=m_J-25-5\log_{10}(d_L)-K ,
\end{equation}
where the luminosity distance is $d_L=d(1+z)$, $d$ is the co-moving
distance in the units of \Mpc, and $z$ is the observed redshift. The
term $K$ is the $k\!+\!e$-correction, adopted according to
\citet{norberg:02}.

\subsection{Selection effects: visibility of galaxies at different distances}

To calculate the LF of galaxies we need to know the number of galaxies
of a given luminosity per unit volume.  The principal selection effect
in flux-limited surveys is the absence of galaxies fainter than the
survey limiting magnitude.  This effect is well seen in
Fig.~\ref{fig:1} that shows the luminosities of the first-ranked
galaxies at various distances from the observer.

To take this effect into account in the determination of the LF of
group galaxies we used the standard $V_\mathrm{max}^{-1}$ weighting
procedure.  The differential luminosity function $n(L)$ (the expectation of
the number density of galaxies of the luminosity $L$) can be found as follows:
\begin{equation}
n(L) {\mathrm d}L = \sum_i\frac{\mathbf{I}_{(L,L+\mathrm{d}L)}(L_i)}
{V_{\mathrm{max}}(L_i)},
\label{vmax}
\end{equation}
where $\mathrm{d}L$ is the luminosity bin width,
$\mathbf{I}_A(x)$ is the indicator function that selects the galaxies 
that belong to a particular luminosity bin, $V_{\mathrm{max}}(L)$ is the
maximum volume where a galaxy of a luminosity $L$ can be observed in the
present survey, and the sum is over all galaxies of the survey.
This procedure is non-parametric, and gives both the form and true 
normalization of the LF.

We select galaxies in the distance interval 70--500\,\Mpc.  At small
distances, bright galaxies are absent from the survey (see
Fig.~\ref{fig:1}) due to the limiting bright apparent magnitude of the
survey.  To avoid this selection effect, we set the lower distance
limit to 70\,\Mpc. The upper limit is set to 500\,\Mpc\ since at large
distances the weights for restoring group luminosities become too big
(see Fig.~\ref{fig:14}).

\subsection{Determination of environmental densities}\label{sec:2.3}

Already early studies of the distribution of galaxies of different
luminosity showed that clustering of galaxies depends on their
luminosity \citep{hamilton:88,einasto:91c}, and thus the LF of
galaxies depends on the environment where the galaxy is located
\citep{cuestabolao:03,mo:04,croton:05,hoyle:05,xia:06}. 
Recent studies have
demonstrated that both the local (group) as well as the global
(supercluster) environments play a role in determining
properties of galaxies, including their luminosities
\citep{einasto:07c}.  To estimate these effects and to investigate the
dependence of the galaxy LF on the environment we calculated the
luminosity density field.

\begin{table}
\caption{The numbers of first-ranked, satellite and isolated galaxies in
different environments}
\label{Tab2}
\begin{tabular}{ccccc}
\hline\hline
\noalign{\smallskip}
Population  & D1    & D2 & D3 & D4 \\
            & Void  & Filament & Supercluster & SC Core \\
            & $D\!\leq1.5$ & $1.5<\!D\!\leq4.6$ & $4.6<\!D\!\leq7$ & $D\!>7$ \\
\hline
\noalign{\smallskip}
First-ranked &  5499 & 11979 & 3511 & 2261 \\
Satellites   &  8196 & 24078 & 9561 & 8582 \\
Isolated gal. & 36813 & 40553 & 8263 & 4359 \\
\hline
\end{tabular}
\begin{list}{}{}
\item The density parameter $D$ is the global environmental density in
  units of the mean density (see Sect.~\ref{sec:2.3} for more
  information).
\end{list}
\end{table}

To calculate the density field we need to know the expected total
luminosities of groups and isolated galaxies (a detailed description
of calculating these luminosities is given in
Sect.~\ref{sec:totlum}).  These quantities are given in the group
catalogue by T06.  The luminosity density fields were found using kernel
smoothing as
described in our 2dFGRS supercluster catalogue \citep{einasto:07b}:
\begin{equation}
\label{D}
D(\mathbf{x})=\sum_i K(\mathbf{x-x}_i;h)\,L_i,
\end{equation}
where $K(\mathbf{x};h)$ is a suitable kernel of a width $h$ with a unit 
volume integral, and $L_i$ is 
the luminosity of the $i$-th galaxy. The sum extends formally over all galaxies,
but
the kernel is usually chosen to differ from zero only in a limited range of the
argument; this limits the number of galaxies in the sum. For details see
\citet{einasto:07b}.

We used the
Epanechnikov kernel: 
\begin{equation}
\label{Epan}
K_{\mathrm E}(\mathbf{x};h)=\frac{15}{8\pi h^3}(1-\mathbf{x}^2/h^2)\;,\quad
\mathbf{x}^2\leq h^2,\;  
\quad 0\quad \mathrm{otherwise},  
\end{equation}
of a radius $h=8\,$\Mpc.

After that, we divided all groups (galaxies) into classes, according to
the value of the global environmental density $D$ at their location 
as follows: low
  density regions with $D \leq 1.5$, medium density regions with $1.5
  < D \leq 4.6$, high density regions with $4.6 < D \leq 7$, and very
  high density regions with $D > 7$ (all densities are in units of the
  mean luminosity density for the sample volume); 
  we denote these regions as D1, D2, D3, D4,
  respectively. The threshold density 4.6 was used in our
supercluster catalogue \citep{einasto:07b} to separate superclusters
from field objects. We define superclusters as non-percolating
  high-density regions of the cosmic web using the global density to
  discriminate between objects belonging to superclusters or to the
  field (see Fig.~\ref{fig:16} below for illustration).
\citet{einasto:07c} showed that the densities $D > 7$ are
characteristic to supercluster cores. High density cores are present
in rich superclusters only. The supercluster environment
  represents poor superclusters and the outskirt regions of
  rich superclusters.  As seen from Fig.~\ref{fig:16}, supercluster
  cores may have a complex internal structure, consisting of clusters
  and groups and even isolated galaxies.  The threshold density 1.5
  that separates the low and medium density regions corresponds
  approximately to the division between galaxies in
  voids and those in filaments.  These divisions
  are not directly related to the shape or other properties of galaxy
  systems (for a recent discussion of the properties of DM haloes in
  different environments see \citet{hahn:07}).

We use these four density classes to study the environmental
dependences of the LFs. In Table~\ref{Tab2} we show the number of
galaxies in different environments for different populations (first-ranked
galaxies, group satellite galaxies and isolated
galaxies). Everywhere in this paper the supercluster class usually
does not include galaxies in supercluster cores; if we lump these
classes together, we tell that.

\section{Luminosity functions in different environments}\label{sec:3}

\subsection{Brightest group galaxies}

\begin{figure}
\resizebox{\hsize}{!}
{\includegraphics{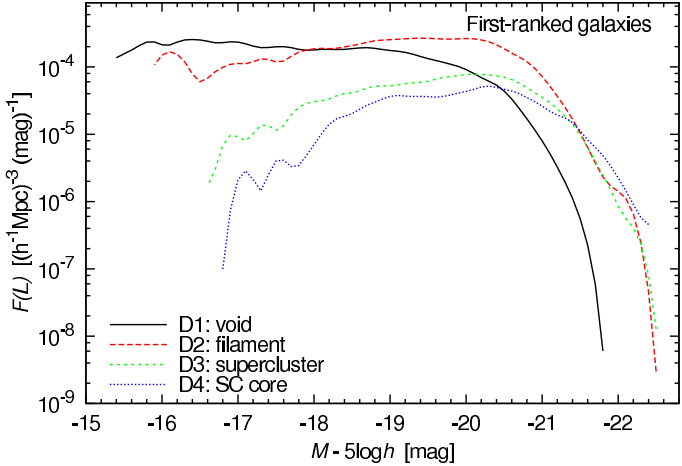}}
\caption{The differential LFs of first-ranked galaxies.  The functions
  have been calculated for four classes of global environmental
  density: D1, D2, D3, D4 (voids, filaments, superclusters, cores
    of superclusters), with limiting densities 1.5, 4.6 and 7 in
  units of the mean density. Solid line -- void galaxies; dashed line
  -- filament galaxies; short dashed line -- supercluster galaxies;
  dotted line -- supercluster core galaxies.}
\label{fig:2}
\end{figure}

We use our galaxy samples and the catalogue of groups of galaxies
(T06) to calculate the LF for first-ranked (brightest group) galaxies.  The
catalogue by
T06 gives for each group the luminosity of the first-ranked galaxy
(most luminous in the $b_J$-filter).  In the present study we shall made
no effort to use for the first-ranked galaxy identification other galaxy
properties, such as spectral type, colour index or possible activity.
These morphological aspects deserve a more detailed study which is
outside the scope of the present investigation.

We calculated the differential LF of first-ranked galaxies in various
environments for different samples. The numbers of galaxies used are
given in Table~\ref{Tab2}.

The differential LFs of first-ranked galaxies are shown in
Fig.~\ref{fig:2} for different environmental densities. In order not
to overcrowd the figures, we do not show error bars. As there are many
galaxies, errors are small; typical errors are illustrated in
Fig.~\ref{fig:4}.

Figure~\ref{fig:2} shows that there exist large differences between
LFs in different environmental regions.  The brightest first-ranked
galaxies in void regions have a factor of 3--4 lower luminosity than
the brightest first-ranked galaxies in regions of higher environmental
densities.  For this reason the whole LF of void first-ranked galaxies
is shifted toward lower luminosities. At the same time, there are no
significant differences between the luminosities of the brightest
first-ranked galaxies in the filament, supercluster, and supercluster
core environments. Later we shall see that the same is valid for
satellite galaxies.

The second large difference between the first-ranked galaxy
luminosities in various environments is the presence of a well-defined
{\em lower} limit of first-ranked galaxy luminosities in the
superclusters and cores of superclusters.  In supercluster cores the
lower first-ranked galaxy luminosity limit is about
$-17\,$mag.  When we move to lower environmental
densities, the lower first-ranked galaxy luminosity limit gets smaller
(see Fig.~\ref{fig:2}). The supercluster core environment forces lower
limits also for other galaxies (satellites and isolated galaxies). The
void and filament first-ranked galaxies do not have any lower
luminosity limit.

\subsection{Group second-ranked galaxies}

We define the group second-ranked galaxy as the most luminous satellite
galaxy in the group: it is the second luminous galaxy in the group.

\begin{figure}
\resizebox{\hsize}{!}
{\includegraphics{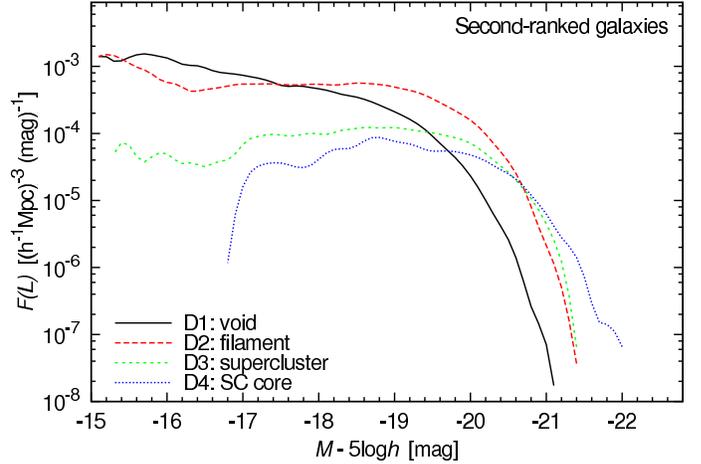}}
\caption{The differential LFs of
  group second-ranked galaxies. Labels are the same as in Fig.~\ref{fig:2}.
}
\label{fig:3}
\end{figure}

\begin{figure}
\resizebox{\hsize}{!}
{\includegraphics{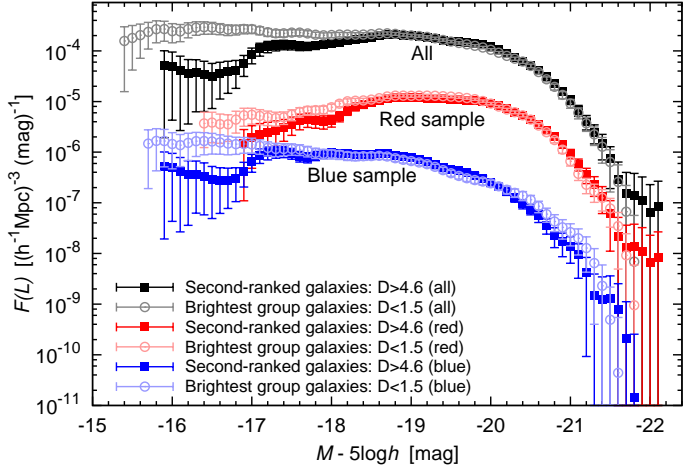}}
\caption{The differential LFs of two populations: filled squares --
second-ranked galaxies
  in the supercluster (including the supercluster core) environment ($D>4.6$);
  open circles -- first-ranked galaxies in the void environment
($D<1.5$). Upper (black) points are for all galaxies; middle (red) points are
for the red population (shifted down by one unit in the $\log(F(L))$ scale);
bottom
(blue) points are for the blue population (shifted down by two units in
the $\log(F(L))$ scale). }
\label{fig:4}
\end{figure}

In Fig.~\ref{fig:3} we plot the differential LFs of group
second-ranked galaxies in different environments.  The overall picture
is similar to the LFs of first-ranked galaxies.  The primary
difference is that the bright end of the LF is shifted to lower
luminosities. The faint-end limits are approximately the same as for
first-ranked galaxies.

Another difference (compared with first-ranked galaxies) is that the
brightest second-ranked galaxies 
in the supercluster core environment are more
luminous than the brightest galaxies in the supercluster and filament
environments; for first-ranked galaxies the bright end of the LFs was
the same for these three environments. 

This effect is expected if second-ranked galaxies in high density environments
had been first-ranked galaxies before they were drawn into a larger
cluster via merging of groups into larger systems.
Thus the LF of second-ranked
galaxies in high density regions is more close to the LF of first-ranked
galaxies than to the LF of second-ranked galaxies in low
density regions.

To test this last assumption, we plotted in Fig.~\ref{fig:4} the
differential LFs of two populations: the first population consists of
the second-ranked galaxies in the supercluster environment (including
the supercluster core environment); the second sample includes the
first-ranked galaxies in the void region. The error-bars in this plot are
Poisson 1-$\sigma$ errors; in other figures (where only lines are
shown) the errors have the same order of magnitude. We see that these
two distributions are pretty similar.  There are differences at the
faint end, but these are caused by the environment; there are only a
few faint first- and second-ranked galaxies in
high-density regions. Thus we see that the second-ranked galaxy LF in
high-density regions is similar to the first-ranked galaxy LF in the
low-density environment.

To show that this similiarity is not accidental, we divided these galaxies into
two samples (red and blue galaxies), using information about the colours of
galaxies (the rest-frame colour index, $col=(B-R)_0$, \citet{cole:05}). We used
the limit $col=1.07$ to separate the populations of red (passive) and blue
(actively star-forming) galaxies. For these two samples the LFs are still
similar (see Fig.~\ref{fig:4}), only the supercluster galaxies are in
average 0.15\,mag redder than the void galaxies.
This shift was also seen in \citet{einasto:07c} where we showed that 
not only the brightest galaxies, but all galaxies in 
superclusters are redder than those in voids.

The first-ranked galaxy population is different from that of the
satellite galaxy population, but the changeover from one population to
the other is smooth.

\subsection{Other satellite galaxies}

\begin{figure}
\resizebox{\hsize}{!}
{\includegraphics{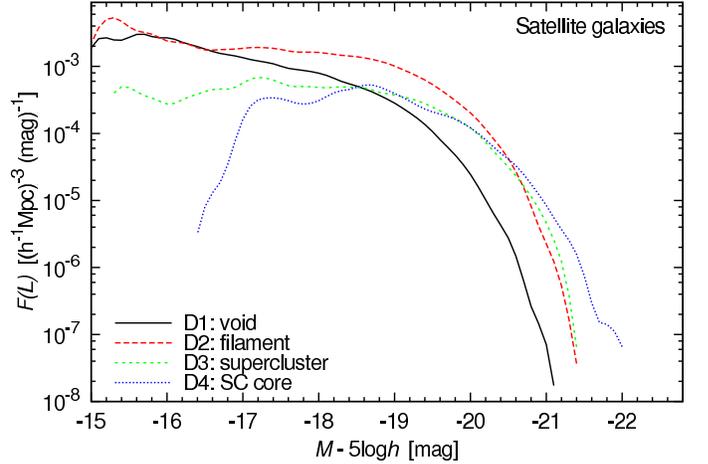}}
\caption{The differential LFs of
  satellite galaxies. Labels are the same as in Fig.~\ref{fig:2}.
}
\label{fig:5}
\end{figure}

Group satellite galaxies are all galaxies in groups (excluding first-ranked
galaxies).  The satellite galaxy population includes also the
second-ranked galaxies.

In Fig.~\ref{fig:5} we show differential LFs of group satellite galaxies.
The LFs are similar to the LFs of second-ranked galaxies.  The primary
difference is that the faint satellites in the supercluster and supercluster 
core environments have
lower luminosities, and there are more faint satellites
than faint second-ranked galaxies. In supercluster
cores, the brightest satellites are more luminous
than those in lower density environments. In addition, 
in the highest density environment there exists a
sharp lower satellite luminosity limit.

\subsection{Isolated galaxies}

\begin{figure}
\resizebox{\hsize}{!}
{\includegraphics{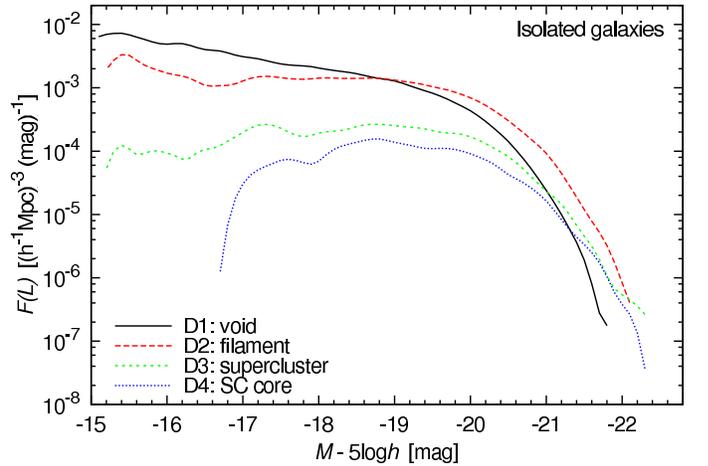}}
\caption{The differential LFs of
  isolated galaxies. Labels are the same as in Fig.~\ref{fig:2}.
}
\label{fig:6}
\end{figure}

We define all galaxies that do not belong to groups/clusters in the 
T06 group catalogue, as isolated galaxies. In this section we present
the LFs of isolated galaxies; we shall discuss the nature of isolated
galaxies later, in a separate section.

In Fig.~\ref{fig:6} we show the differential LFs of isolated galaxies
in different environments. Isolated galaxies in the supercluster core
environment have a faint luminosity limit (similar to other populations).
The differences in the LFs of bright isolated galaxies
between various environments are much
smaller than for other populations. The brightest isolated galaxies occur in
the filament environment (for other populations the brightest galaxies can
be found in the supercluster core environment). 
This means that in the supercluster and supercluster core
environments the brightest galaxies lie in groups/clusters,
contrary to the void and filament environments, where many bright galaxies are
identified as isolated galaxies.

\begin{figure*}
\centering
\includegraphics[width=18cm]{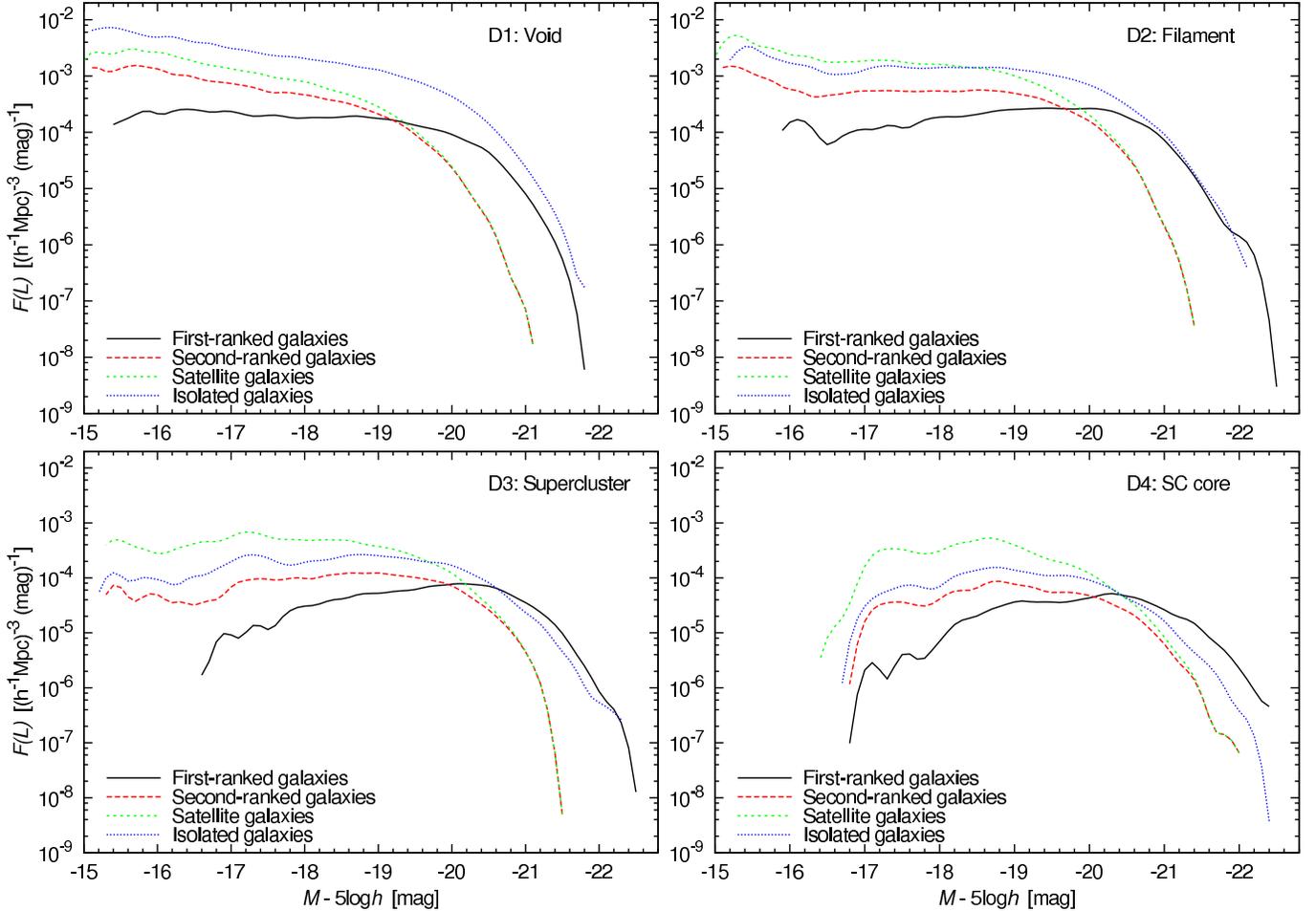}
\caption{Differential LFs in different environments and for different
  galaxy populations. Top-left panel -- void environment; top-right
  panel -- filament environment; bottom-left panel -- supercluster
  environment; bottom-right panel -- supercluster core
  environment. Solid line shows first-ranked galaxies; dashed line --
  second-ranked galaxies; short-dashed line -- satellite galaxies;
  dotted line -- isolated galaxies.}
\label{fig:7}
\end{figure*}

\subsection{Comparison of  LFs in different environments}

Differential LFs of galaxy populations in various environments are
shown in Fig.~\ref{fig:7}. Each panel represents a different
environment, and in each panel we show the LFs of different
populations (first-ranked, second-ranked, satellite and isolated
galaxies).

Figure~\ref{fig:7} (upper-left panel) shows that in voids, the bright
end of LFs of all galaxy populations is shifted toward lower
luminosities -- in the void environment first-ranked galaxies of
groups are fainter than those in higher density environments. We
noticed this effect also in \citet{einasto:07c}. Interestingly, the
bright end of the LF for isolated galaxies in voids is comparable to
that of first-ranked galaxies.  We discuss the possible reasons for
that in the next section. The LFs of second-ranked galaxies and of all
satellites are comparable, although there is a slight increase of the
LF of satellite galaxies at the lowest luminosities. The LF of first-ranked
galaxies, in contrary, has a plateau at the faint end, without
signs of increase.

In the filament environment (Fig.~\ref{fig:7}, upper-right panel) the
bright ends of the LFs for first-ranked galaxies and for isolated
galaxies are similar, while the brightest second-ranked galaxies and
satellite galaxies are fainter. For a wide range of luminosities, the
LF for first-ranked galaxies is slowly decreasing toward fainter
luminosities, while LFs for other galaxy populations have a plateau.

The LFs for the supercluster environment (excluding supercluster
cores) are shown in the lower-left panel of Fig.~\ref{fig:7}. As we
mentioned, the supercluster environment represents poor superclusters
and the outskirt regions of rich superclusters.  This Figure shows
that the first-ranked galaxies in the supercluster environment have
luminosities comparable to those of the first-ranked galaxies in
filaments, but the LFs for faint galaxies differ. Instead of a
plateau, there the LFs show a decrease toward the faint end. The LF
for first-ranked galaxies in this region has a well-defined faint
luminosity limit (approximately $-17\,$mag), while
the LFs for other galaxy populations extend to fainter luminosities.

The LFs for supercluster cores are shown in Fig.~\ref{fig:7},
lower-right panel.  We notice the striking difference between the LFs
in supercluster cores and the LFs in other environments: here all LFs
have a well-defined lower luminosity limit, about
$-17\,$mag, which for first-ranked galaxies was seen
already in supercluster environment. Also, in supercluster cores the
brightest first-ranked galaxies are more luminous than the brightest
first-ranked galaxies in other environments.

Our earlier studies have shown that the most luminous groups are located
in superclusters \citep{einasto:03a,einasto:03c}. Here we see the
same trend for the brightest first-ranked galaxies.

In summary, the most dense environment (supercluster cores) is
different from other environments: there are no very faint galaxies,
and the brightest first-ranked galaxies are brighter than the first-ranked
galaxies in lower density environments. The lower luminosity
limit is shifted to smaller luminosities, if we move to less dense
environments. The transition between different environments is smooth.

\section{Nature of isolated galaxies}\label{sec:4}

We assume that some fraction of isolated galaxies are first-ranked
galaxies of groups/clusters, which have all its fainter members
outside the visibility window of the survey.  The best way to verify
this assumption is actual observation of fainter galaxies around
isolated galaxies; this would need a dedicated observational program.
However, we can check if the presence of fainter companions is
compatible with other data on the distribution of magnitudes of
galaxies in groups.  First, we analyse the LF of isolated galaxies and
examine how many isolated galaxies could actually be first-ranked
galaxies and how this ratio depends on the environment.

\subsection{The luminosity function for isolated galaxies}

The overall shape of the LFs in Fig.~\ref{fig:7} suggests that
isolated galaxies may be a superposition of two populations: the
bright end of their LF is close to that of the first-ranked galaxy LF,
and the faint end of the LF is similar to the LF of satellite
galaxies. This is compatible with the assumption that the brightest
isolated galaxies in a sample are actually the brightest galaxies of
invisible groups.

In the supercluster core environment the brightest isolated galaxies
are fainter than the brightest first-ranked galaxy, but they are
almost as bright as the second-ranked galaxies in this environment.
Earlier we showed that the second-ranked galaxies in high-density
regions are similar to first-ranked galaxies in lower-density regions.
In other words, second-ranked galaxies in supercluster core clusters
can be considered as first-ranked galaxies of clusters before the last
merger event.

In the void environment, the faintest isolated galaxies are brighter than
the faintest galaxies of other populations. This suggests that some
isolated galaxies in voids are truly isolated; they do not
belong (or have belonged) to any groups. Truly isolated galaxies
are rare in denser environments.

\subsection{Magnitude differences between the first-ranked and the
  second-ranked group galaxies}

The simplest test to examine the assumption that isolated galaxies can
be first-ranked galaxies, is the following.  A group has only one
galaxy in the visibility window, if its second-ranked galaxy (and all
fainter group galaxies) are fainter than the faint limit of the
luminosity window at the distance of the galaxy.  Thus we calculated
for each isolated galaxy the magnitude difference $M_\mathrm{diff,iso}
= M_l - M_{b_J}$, where $M_l$ is the absolute magnitude corresponding
to the faint limit of the apparent magnitude window $m_l = 19.45$, and
$M_{b_J}$ is the absolute magnitude of the galaxy in the
$b_J$-filter. These magnitudes should also be corrected for the
$k\!+\!e$-effect, but as the correction is the same for both, it does
not influence their difference.

\begin{figure}
\resizebox{\hsize}{!}
{\includegraphics{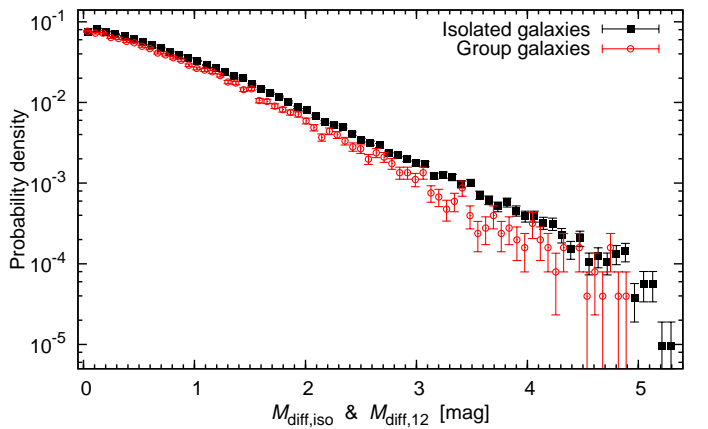}}
\caption{The differential distributions of magnitude differences for
  group first-ranked and second-ranked galaxies (open circles) and of
  magnitude differences between isolated galaxies and the faint limit
  of the visibility window (filled squares).}
\label{fig:8}
\end{figure}

The distribution of magnitude differences should be compared with the
distribution of the actual magnitude differences between the
first-ranked and second-ranked group galaxies, $M_\mathrm{diff,12} =
M_2 - M_1$.  The differential distributions of magnitude differences
between the first-ranked and second-ranked group galaxies
$M_\mathrm{diff,12}$, and the difference $M_\mathrm{diff,iso} = M_1 -
M_{b_J}$ of isolated galaxies, are shown in Fig.~\ref{fig:8}.  The
distributions look rather similar. The main difference is that there
are less very small magnitude differences $M_\mathrm{diff,iso}$ for
isolated galaxies.  In the case of very small magnitude differences
between the first-ranked and second-ranked galaxies the second-ranked
galaxy is also observed for redshifts, and the galaxies are not
isolated.

The overall similarity of both distributions suggests that our assumption
(that isolated galaxies are actually the first-ranked galaxies with fainter
companions located outside the observational window) passes the magnitude
difference test.  Of course,
this test does not exclude the possibility of existence
of truly isolated galaxies.

\subsection{Group visibility at different distances}

As a further test to check our hypothesis concerning the nature of
isolated galaxies we check how well actual nearby groups
are visible, if shifted to larger distances.  For this purpose we
selected two subsamples of groups at different true distances from
the observer (and with different mean absolute magnitude of the first-ranked 
group galaxy, $M_1$).  The first subsample was chosen in a
nearby region with distances $100 \leq d < 200$\,\Mpc, and the number
of visible galaxies $N_\mathrm{gal} \geq 10$.  The other group
sample was chosen in the distance interval $200 \leq d < 300$\,\Mpc, and
$N_\mathrm{gal} \geq 10$.

\begin{figure}
 \resizebox{\hsize}{!}
{\includegraphics{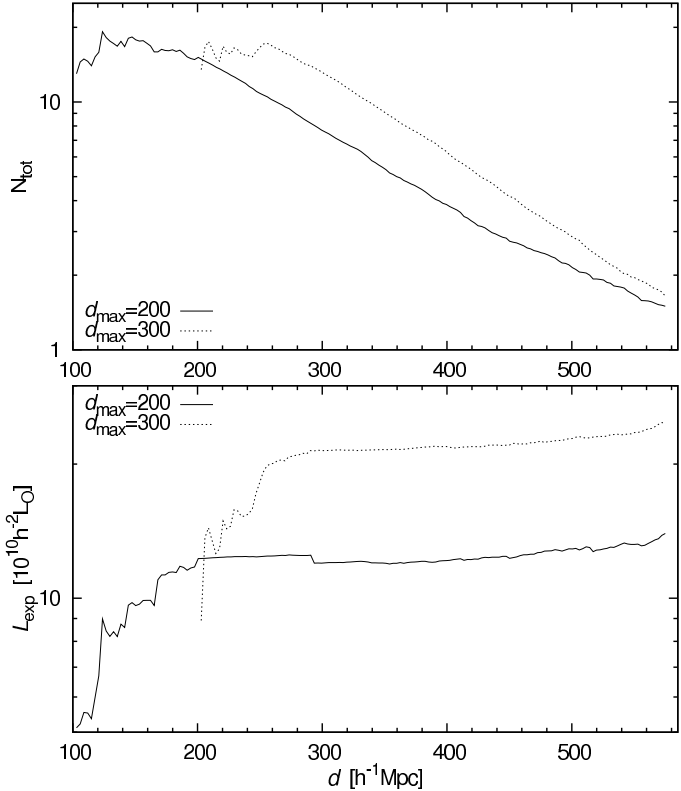}}
\caption{ Upper panel: the mean number of galaxies in shifted groups as a
  function of distance.  Lower panel: restored
  mean total luminosities of shifted groups.
   The solid line shows the results for groups located initially at
  distances $100 \leq d < 200$\,\Mpc; dashed line -- for groups of initial
  distances $200 \leq d < 300$\,\Mpc.}
\label{fig:9}
\end{figure}

\begin{figure}
\resizebox{\hsize}{!}
{\includegraphics{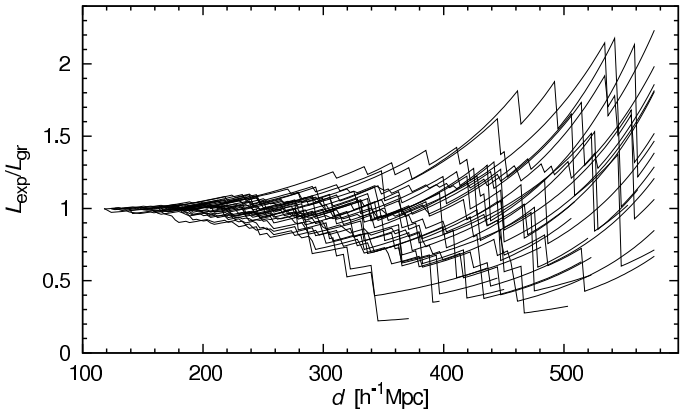}}
\caption{The restored luminosities of shifted groups, in units of the actual
  luminosity at the true distance. 
}
\label{fig:10}
\end{figure}

Next groups were shifted to progressively larger distances, galaxy apparent
magnitudes were calculated, and galaxies inside the visibility window were
selected.  Details of this procedure were described by T06.  The number of
galaxies inside the visibility window for shifted groups decreases; the mean
number of galaxies in shifted groups is shown in the upper panel of
Fig.~\ref{fig:9}.  We see that the mean number decreases almost linearly in
the $\log N$ - $d$ diagram; at the far side of our survey the mean number of
remaining galaxies in groups is between 1 and 2.

The expected total luminosity of groups, calculated on the basis of
galaxies inside the visibility window and using the procedure outlined
in Sect. \ref{sec:totlum} below, is shown on the lower panel of
Fig.~\ref{fig:9}.  We see that the mean values of restored total
luminosity of groups are almost identical with the true luminosity
at the initial distance. At the very far end, the expected total
luminosities of groups are a little higher than the initial
luminosity, i.e. the expected luminosities are slightly over-corrected.
Lower mean luminosities at low distances are caused by the lower
number of groups in this region.   The restored luminosities of individual
groups have a scatter that increases with distance, as seen
from Fig.~\ref{fig:10}. Luminosities in Fig.~\ref{fig:10} are in units
of the actual luminosity at the true distance. We plotted in this
Figure the expected total luminosities for 30 groups, selected in
the region 100--200\,\Mpc, as a function of shifted distance.  This
scatter can be used to estimate errors of estimated total luminosities
of groups as a function of the number of remaining galaxies in
groups.

\subsection{Luminosity functions of brightest+isolated galaxies}

\begin{figure}
\resizebox{\hsize}{!}
{\includegraphics{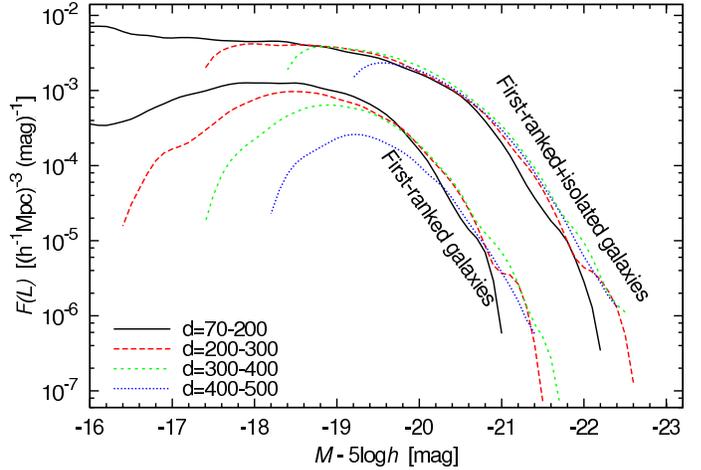}}
\caption{The LFs of the first-ranked and first-ranked+isolated galaxies in
  different distance intervals (distances are in units of \Mpc).  The
  LFs of first-ranked galaxies are shifted to left by 1\,mag.}
\label{fig:11}
\end{figure}

\begin{figure}
\resizebox{\hsize}{!}
{\includegraphics{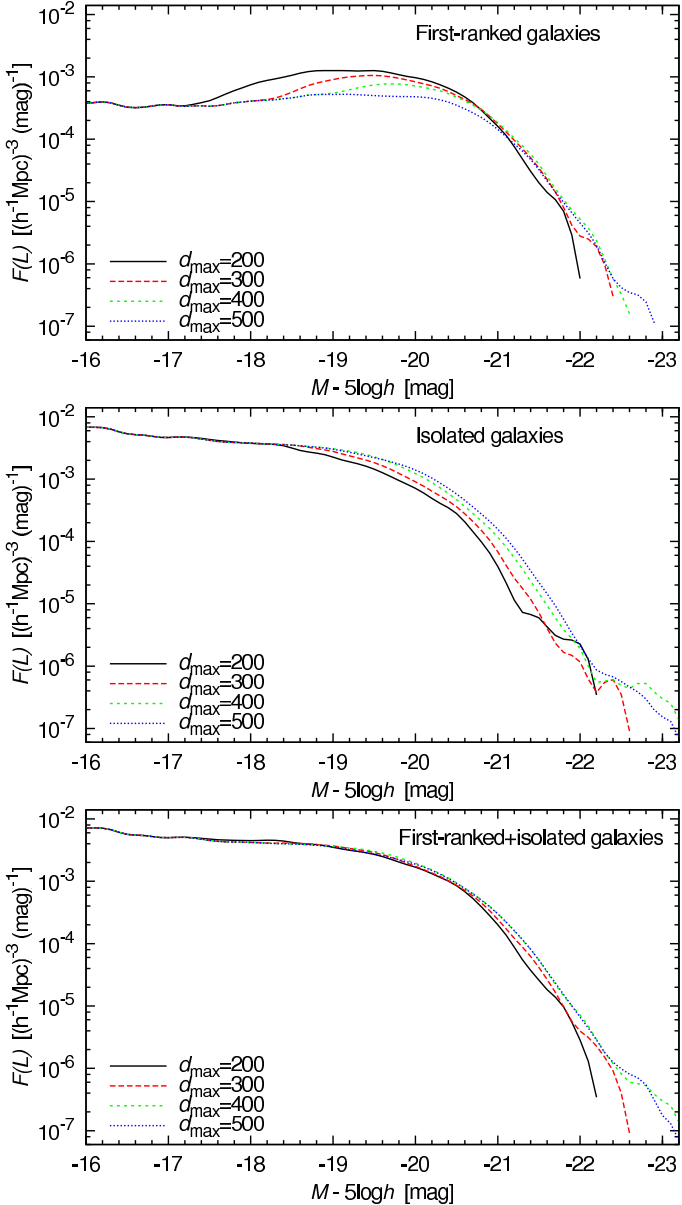}}
\caption{The differential LFs of first-ranked (upper panel),
  isolated (middle panel) and first-raked+isolated (lower panel)
  galaxies for samples of different maximum distances
  ($d_{\mathrm{max}}= 200,300,400,500$\,\Mpc).  }
\label{fig:12}
\end{figure}

To test the assumption that isolated galaxies are first-ranked
galaxies, we can also examine how distance-dependent selection effects
influence the LFs of first-ranked galaxies and isolated galaxies.

\begin{table}
\caption{Numbers of galaxies in distance-dependent samples}
\label{Tab3}
\begin{tabular}{cccc}
\hline\hline
\noalign{\smallskip}
  Distance interval$^{a}$ & First-ranked gal. & Isolated gal. & Fraction$^{b}$
\\ 
\hline
\noalign{\smallskip}
  70--200 & 5184 &16724 & 0.24   \\
 200--300 & 6931 &22928 & 0.23   \\
 300--400 & 7450 &28930 & 0.20   \\
 400--500 & 3888 &22426 & 0.15   \\
\hline
\end{tabular}
\begin{description}{}{}
\item[$^a$] Distances are in units \Mpc.
\item[$^b$] Fraction of first-ranked galaxies in the first-ranked+isolated
sample. 
\end{description}
\end{table}

Figure~\ref{fig:11} shows the LFs of first-ranked and first-ranked+isolated
galaxies in different distance intervals.  The numbers of
first-ranked and isolated galaxies in each distance interval are given
in Table~\ref{Tab3}.  The LFs of first-ranked galaxies are
distance-dependent: with increasing distance the number of faint
galaxies decreases. If we add the isolated galaxy sample (we assume
that isolated galaxies are first-ranked galaxies) to the first-ranked
sample, then the combined LFs are almost independent of distance.  The
remaining differences are only in the lowest luminosity ranges where
data are incomplete; the differences are much smaller than for the
first-ranked samples.

In the second test, we calculated the LFs of first-ranked galaxy, isolated and
first-ranked+isolated galaxies for a number of limiting distances from the
observer: $d_{\mathrm{max}} = 200$,~300,~400 and 500\,\Mpc. The
minimum distance is the same for all samples (70\,\Mpc).  The total
number of first-ranked galaxies in these subsamples is 5184, 12115,
19565 and 23453, respectively.

The calculated LFs are shown in Fig.~\ref{fig:12}. If we look only at
the first-ranked  or the isolated galaxy samples, then the LFs depend on
distance. If we combine these two samples, then the combined LFs are
independent of distance. This supports our assumption that most of
isolated galaxies are actually the first-ranked galaxies with
satellite galaxies outside the visibility window. With increasing
distance, the fraction of (visible) brightest galaxies decreases (see
Table~\ref{Tab3}).  With increasing environmental density, the
fraction of first-ranked galaxies increases (see Table~\ref{Tab2}).

Our tests show that all (or almost all) bright isolated galaxies are actually
first-ranked galaxies. We cannot say that for fainter galaxies: there might be
some fainter galaxies that are truly isolated.
\begin{figure*}
\centering
\includegraphics[width=18cm]{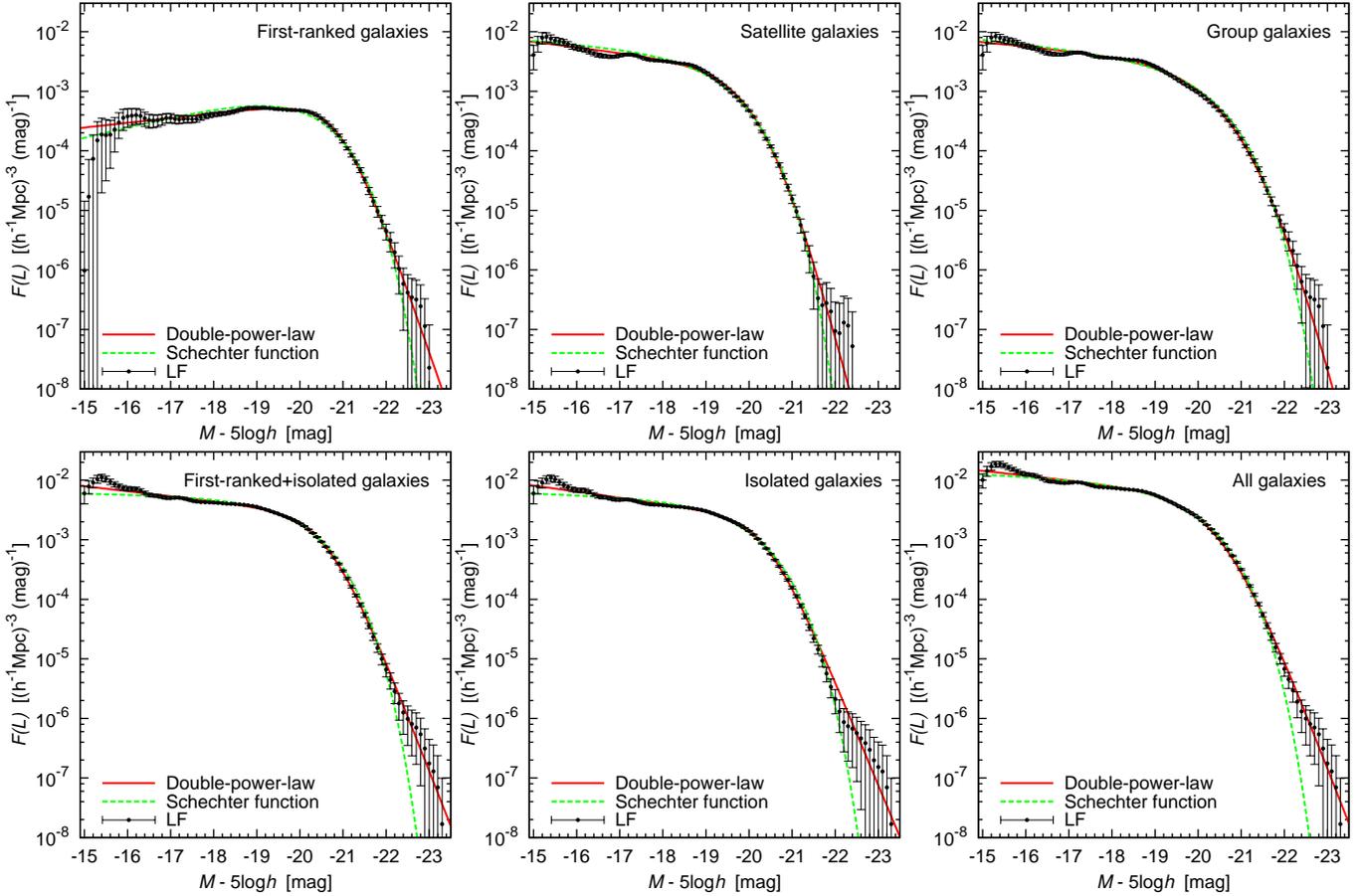}
\caption{Differential LFs for various galaxy populations: first-ranked,
   satellite, first-ranked+satellite (group), first-ranked+isolated, isolated
and all galaxies. The points are LFs, using 2dF
  galaxy catalogue. Error-bars are Poisson 1-$\sigma$ errors. The red solid
  line is the double-power-law and the green dashed line is the
  Schechter function.}
\label{fig:13}
\end{figure*}

\section{Full luminosity functions}\label{sec:5}

\subsection{Comparison of the Schechter and the double-power-law
  luminosity functions}\label{sec:schechter}

A double-power-law form of the group LF was found already by
\citet{christensen:75,kiang:76,abell:77,mottmann:77}.  In these papers
a sharp transition between two power indices at a characteristic
luminosity $L^{*}$ was applied.  We shall use a smooth transition:
\begin{equation}
F (L) \mathrm{d}L \propto (L/L^{*})^\alpha (1 +
(L/L^{*})^\gamma)^{(\delta-\alpha)/\gamma}
\mathrm{d}(L/L^{*}), 
\label{eq:abell}
\end{equation}
where  $\alpha$ is the exponent at low luminosities $(L/L^{*}) \ll 1$,
$\delta$ is the exponent at high luminosities    $(L/L^{*}) \gg 1$,
$\gamma$ is a parameter that determines the speed of transition between the
two power laws, and $L^{*}$ is the characteristic luminosity of the
transition, similar to the characteristic luminosity of the Schechter function.
A similar double-power law was also used by \citet{vale:04} to fit the
  mass-luminosity relation in their subhalo model, and by
  \citet{cooray:05a} to fit the luminosity function of central galaxies.

We shall compare the double-power-law function with the 
popular \citet{schechter:76} function:
\begin{equation}
F (L) \mathrm{d}L \propto (L/L^{*})^\alpha \exp
{(-L/L^{*})}\mathrm{d}(L/L^{*}),
\label{eq:schechter1}
\end{equation}
where $\alpha $ and $L^{*}$ (or the respective absolute magnitude $M^{*}$)
are parameters.

\begin{table*}
\caption{Schechter and double-power-law parameters}
\label{Tab4}
\begin{tabular}{lcc|cccc}
\hline
\noalign{\smallskip}
         & \multicolumn{2}{c|}{Schechter} &
\multicolumn{4}{c}{Double-power-law} 
\\
\noalign{\smallskip}
  Sample & $\alpha$ & $M^*$ &
  $\alpha$ & $\gamma$ & $\delta$ & $M^*$ \\ 
\hline
\noalign{\smallskip}
  First-ranked galaxies& $-0.557\pm 0.031$ & $-19.95\pm 0.04$ & 
  $-0.802\pm 0.015$ & $2.44\pm 0.13$       & $-6.22\pm 0.40$ & $-20.97\pm 0.07$
\\
  Satellite galaxies   & $-1.131\pm 0.026$ & $-19.18\pm 0.04$ & 
  $-1.231\pm 0.031$ & $1.73\pm 0.19$       & $-7.95\pm 1.84$  & $-20.44\pm 0.31$
\\
  Group galaxies       & $-1.197\pm 0.020$ & $-19.96\pm 0.04$ & 
  $-1.177\pm 0.044$  & $1.23\pm 0.19$      & $-8.43\pm 3.53$  & $-21.44\pm 0.72$
\\
  First-ranked+Isolated& $-1.031\pm 0.016$ & $-19.93\pm 0.03$ & 
  $-1.211\pm 0.013$ & $2.14\pm 0.14$   & $-5.64\pm 0.48$ & $-20.69\pm 0.11$ \\
  Isolated galaxies    & $-1.058\pm 0.017$ & $-19.76\pm 0.03$ & 
  $-1.258\pm 0.013$ & $2.37\pm 0.17$   & $-5.35\pm 0.41$ & $-20.43\pm 0.10$   \\
  All galaxies         & $-1.098\pm 0.012$ & $-19.77\pm 0.02$ & 
  $-1.200\pm 0.022$  & $1.66\pm 0.15$   & $-5.71\pm 0.71$ & $-20.60\pm 0.20$ \\
  Groups               & $-0.928\pm 0.020$ & $-22.05\pm 0.04$   & 
  $-0.734\pm 0.045$  & $0.89\pm 0.10$   & $-5.98\pm 0.76$ & $-23.33\pm 0.60$ \\
\hline
\end{tabular}

\medskip
$M^*$ is in units of mag$-5\log(h)$.
\end{table*}

Figure~\ref{fig:13} presents the LFs for various galaxy populations:
first-ranked , satellite, first-ranked+satellite (group), first-ranked+isolated,
isolated and all galaxies. When calculating LFs, we
have selected galaxies from all density regions.  The LFs have a
well-defined bend around $L^* \simeq 10^{10}\,h^{-2}\mathrm{L}_{\sun}$
($M^*-5\log h \simeq -20$), and an almost constant level for
luminosities $M^*-5\log h > -19$: the LFs slightly increase by moving
toward lower luminosities, except for the first-ranked galaxy sample,
where the LF is decreasing.

For all galaxy populations we have fitted the Schechter and double-power-law
functions for these LFs. The Schechter and double-power-law parameters with
error
estimates for each sample are given in Table~\ref{Tab4}.  In general, both
functions give a pretty good fit.  Since the double-power-law has more
free parameters, the fit is slightly better. There is still one big
difference
between the Schechter and the double-power-law: for most populations, 
the Schechter
law predicts too few bright galaxies; the double-power-law gives a much better
fit for the bright end of the LF and a better fit in the bend region.

\subsection{Determination of expected total
  luminosities of groups } \label{sec:totlum}

The main problem in the calculation of the group LF is the reduction
of observed group luminosities to expected total luminosities which
take into account galaxies outside the visibility window.  The 2dFGRS
is a flux-limited survey, since very bright as well as faint galaxies
cannot be observed for redshifts using the multifibre technique.  The
estimated total luminosity by T06 was found using the
\citet{schechter:76} LF of galaxies, as done also by
\citet{moore:93,tucker:00}.

In calculating the luminosities of groups we regard every galaxy as
a visible member of a density enhancement within the visible range of
absolute magnitudes, $M_1$ and $M_2$, corresponding 
to the observational window of apparent magnitudes, $m_1$
and $m_2$, at the distance of the galaxy. 
This assumption is based on observations of nearby
galaxies, which indicate that practically all galaxies are located in
systems of galaxies of various size and richness. In this paper we
came to similar conclusions that truly isolated galaxies are rare,
and most observed isolated galaxies are actually the first-ranked galaxies.

To estimate the expected total luminosity of groups we assume that
the LFs derived for a representative volume can be applied also for
individual groups and galaxies.  Under this assumption the estimated
total luminosity per one visible galaxy is
\begin{equation}
L_\mathrm{tot} = L_\mathrm{obs} W_L, 
\label{eq:ldens}
\end{equation}
where $L_\mathrm{obs}=\mathrm{L}_{\sun}10^{0.4\times (M_{\sun}-M)}$ is the
luminosity of the visible galaxy of absolute magnitude $M$ (in units
of the luminosity of the Sun, $\mathrm{L}_{\sun}$), and 
\begin{equation}
W_L =  {\frac{\int_0^\infty L\,F
(L)\mathrm{d}L}{\int_{L_1}^{L_2} L\,F(L)\mathrm{d}L}} 
\label{eq:weight2}
\end{equation}
is the luminous-density weight (the ratio of the expected total luminosity to
the expected luminosity in the visibility window). $L_1$ and $L_2$ are lower and
upper limit of the luminosity window, respectively.  In our calculations we
have adopted the absolute magnitude of the Sun in the $b_J$ filter $M_{\sun} =
5.33$ \citep{eke:04b}.  Further we have adopted the
$k\!+\!e$-correction according to \citet{norberg:02}.

\begin{figure}
\resizebox{\hsize}{!}
{\includegraphics{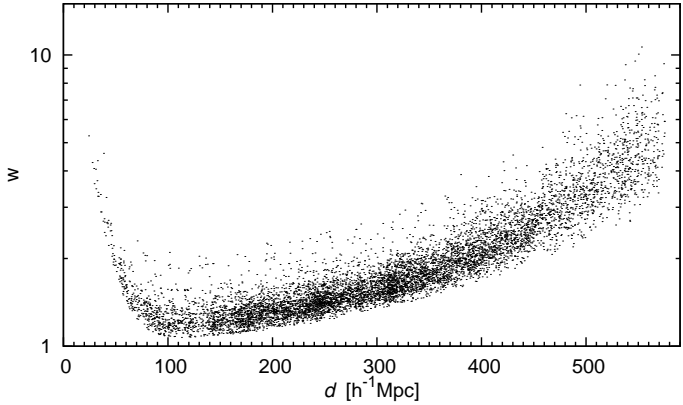}}
\caption{The weights used to correct for
  invisible galaxies outside the observational luminosity window. 
}
\label{fig:14}
\end{figure}

In T06 paper we used the Schechter function for calculating the weights.
In this paper we use the double-power-law instead of the Schechter one,
as the double-power-law represents better the bright end of the LFs. For
weights, we use the double-power-law derived from the full galaxy
sample.  The weights assigned to galaxies as a function of distance from
the observer are shown in Fig.~\ref{fig:14}.  At a distance $d
\approx 100$\,\Mpc\ weights are close to unity; here the observational
window of apparent magnitudes covers the absolute magnitude range of
the majority of galaxies.  Weights rise toward very small distances
due to the influence of bright galaxies outside the observational
window, which are not numerous but are very luminous.  At larger
distances the weight $W_L$ rises again due to the influence of faint
galaxies outside the observational window.  The fairly large scatter
of weights at any given distance is due to differences of the
$k\!+\!e$-correction for galaxies of various energy distribution
parameter $\eta$, and the scatter of the incompleteness correction.

\begin{figure}
\resizebox{\hsize}{!}
{\includegraphics{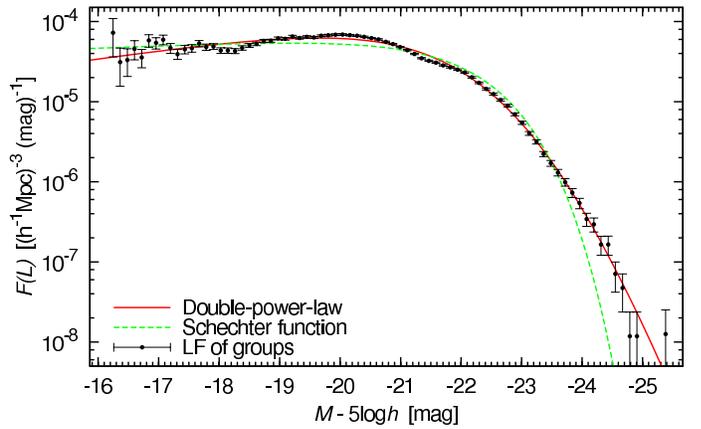}}
\caption{The differential LF of groups (shown by points). The red
  solid line is the double-power-law fit and the green dashed line is
  the Schechter function.  }
\label{fig:15}
\end{figure}

As we have the total luminosities of groups, we are able to calculate
the group LF.  It is plotted in Fig.~\ref{fig:15}. Compared with the
galaxy LF, the turn-off is shifted toward brighter luminosities and
the LF is shallower. Here again, the Schechter function predicts too
few bright groups. The Schechter and double-power-law parameters for
the group LF are given in Table~\ref{Tab4}.

\section{Discussion}\label{sec:6}

\subsection{Distribution of  groups in various environments}

\begin{figure}
\resizebox{\hsize}{!}
{\includegraphics{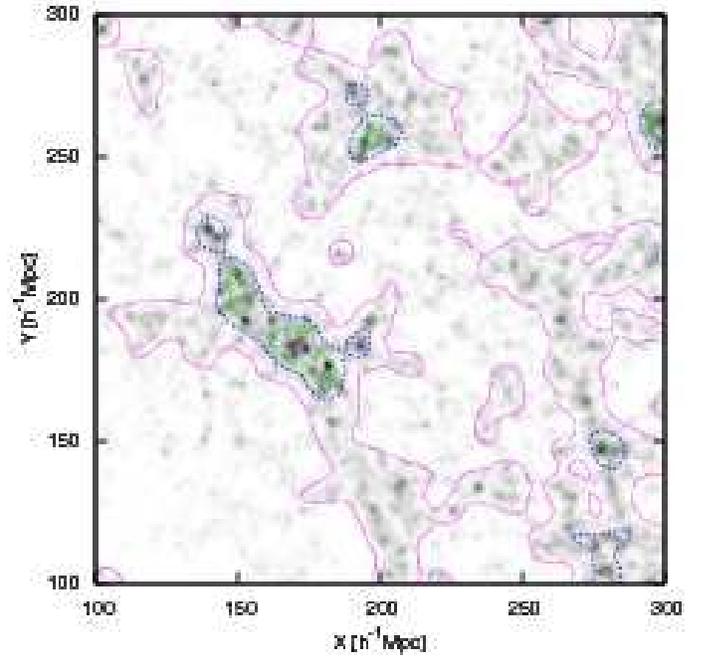}}
\caption{The projected luminosity density distribution of the
    central area of the 2dFGRS Northern sample. The green long-dashed line
    surrounds the supercluster core region, the blue dashed line -- the
supercluster
    region, and the dotted violet line -- the filament region, the rest is the
void
    region. The luminous supercluster near the center of the Figure is
    SCL126, according to the catalogue by \citet{einasto:97a}.}
\label{fig:16}
\end{figure}

To visualise the environmental dependence of the LFs we show in
Fig.~\ref{fig:16} a 2-D luminosity density distribution in
  different regions of the global density. 
  We show a slice of the 2dFGRS with a
  thickness of 100\,\Mpc~ at a distance $d=250$\,\Mpc, perpendicular
  to the line-of-sight, with suitably chosen Cartesian coordinates within
  the plane of the slice. We see that
  in supercluster and their core regions there are numerous dense
  knots -- rich clusters of galaxies.  Knots (clusters/groups) in
  filament regions form elongated clouds around superclusters; often
  these clouds continue filaments inside superclusters.  The most
  luminous supercluster seen in this Figure is SCL126.  It is
  connected with neighbouring superclusters by numerous filaments.

\subsection{Evolution of groups in various environments}
 
To understand the differences between group LFs we performed an
analysis of evolution of haloes (simulated groups) in different
global environments.  For this purpose we simulated the evolution of a
model universe with standard cosmological parameters
$\Omega_\mathrm{m} = 0.27$, $\Omega_{\Lambda} = 0.73$, $\sigma_8 =
0.84$, $z_\mathrm{init} = 500$, $N_\mathrm{part} = N_\mathrm{grid} =
256^3$ in a box of size $L = 256$\,\Mpc.  Particle positions and
velocities were stored for epochs $z = 100$, 50, 20, 10, 5, 2, 1, 0.5
and 0. The density field was calculated for all epochs using two
smoothing kernels, the Epanechnikov kernel of the radius of 8\,\Mpc,
and the Gaussian kernel of the rms width of 0.8\,\Mpc.  These fields
define the global and local environment, respectively.  For the
present epoch the high-resolution density field was used to find
compact haloes. The haloes were defined as all particles within a
box $\pm 2$ cells around the cell of the peak local density. All
together 41060 haloes were found. For each halo its position, peak
local density, global environmental density at its location, and mass
were stored; the mass was found by counting the number of particles in
groups (the mass of each particle is $7.487 \times
10^{10}\,\mathrm{M}_{\sun}$).  For all haloes particle identification
numbers were stored, so it was easy to find positions of particles of
present-epoch haloes at earlier epochs.

\begin{figure}[ht]
\resizebox{\hsize}{!}
{\includegraphics{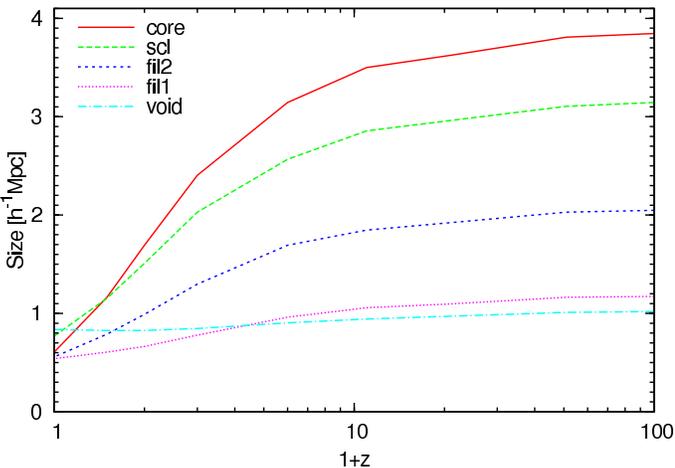}}
\caption{The evolution of the size of samples of particles, 
  collected in haloes at the present epoch, as a function of redshift $z$.
  Present time haloes have been selected in regions of different global
  density at the present epoch, corresponding to supercluster cores,
  superclusters, filaments of various global density, and in voids. 
}
\label{fig:17}
\end{figure}

 To follow changes of the size of the clouds of particles in present-day
 haloes we calculated the mean sizes of particle clouds for earlier
epochs.
 In the present study we are not interested in the changes of positions of
 groups during their evolution, thus we found the mean position of each
 halo and its sizes along three coordinate directions;
 the mean of these sizes was taken as the size of the
 halo.  Halo samples for this study were collected in 5 global
density
 regions at the present epoch, which correspond to cores of superclusters,
 superclusters, rich and poor filaments, and voids.  This division
 corresponds approximately to observed group populations located in various
 global density environments, studied above.

The changes of mean sizes of model haloes located in various global
environment are presented in Fig.~\ref{fig:17}.  The Figure shows dramatic
differences in the evolution of halo sizes. Haloes in void regions
have
almost identical sizes over the whole time interval used in this study.  In
contrast, the sizes of haloes in present core regions of superclusters
were
much larger at earlier epochs: the sizes have decreased by a factor of about
5.  In regions of intermediate global density the changes are the smaller, the
lower is the environmental density.

These differences in the evolution of halo mean sizes are mainly
due to differences in the merger history of haloes in various
environments.  In high-density regions the present-day haloes are
collected from numerous smaller haloes formed independently
around the present brightest group.  During this process the mass of
the halo increases.  The merger rate is a function of the
environmental density, thus we observe gradual changes of the LFs of
first-ranked galaxies.

\begin{figure}[ht]
\resizebox{\hsize}{!}
{\includegraphics{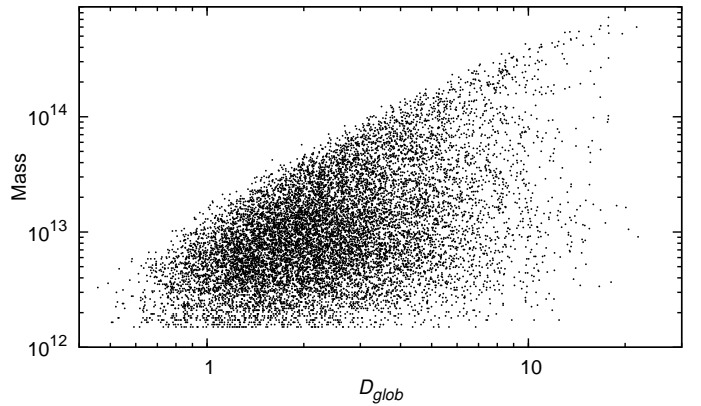}}
\caption{The masses of haloes in regions of different global
  density at the present epoch.  Haloes have been found by the FoF
  method using the same model as in Fig.~\ref{fig:17}.}
\label{fig:18}
\end{figure}

The masses of haloes at the present epoch as a function of the
  global environmental density are shown in Fig.~\ref{fig:18}. For a
  given environmental density the masses have a well-defined upper
  limit, which increases over two orders of magnitude when we move
  from the void environment to the supercluster core environment.  As argued by
  \citet{einasto:94a}, this has a simple explanation.  In all regions where
the density is below average, the density decreases continuously and
there is no possibility to form compact objects -- galaxies.  In
regions where the density is above the average, it increases until
matter collapses to form haloes. The growth of density
is the more rapid the higher is the density. This is the usual gravitational
instability process; for a recent simulation of
that see, e.g., \citet{gao:05}.

Similar halo evolution histories in various environments 
were recently constructed by  \citet{romeo:08},
using  detailed hydrodynamical simulations. Empirical evidence
for differences in galaxy evolution comes among other sources also from the
Millennium Galaxy Catalogue, see \citet{driver:06}.

\subsection{Comparison with earlier studies}

The LF of group brightest and satellite galaxies was investigated
  recently by a number of authors.  \citet{yang:08a} used the Data
Release 4 of the Sloan Digital Sky Survey to study the LF in the
framework of a long series of papers devoted to the halo occupation
distribution \citep[][and references in
\citet{yang:08a}]{vandenbosch:03,yang:03,yang:04,vandenbosch:04,vandenbosch:05,
  zheng:05, yang:05b,vandenbosch:08}. Similar studies were made
  also by \citet{tinker:05,cooray:05b,cooray:05a, milosavljevic:06,
    tinker:06, vale:06,tinker:07, hansen:07, tinker:08, vale:08}.

\begin{figure}[ht]
\resizebox{\hsize}{!}
{\includegraphics{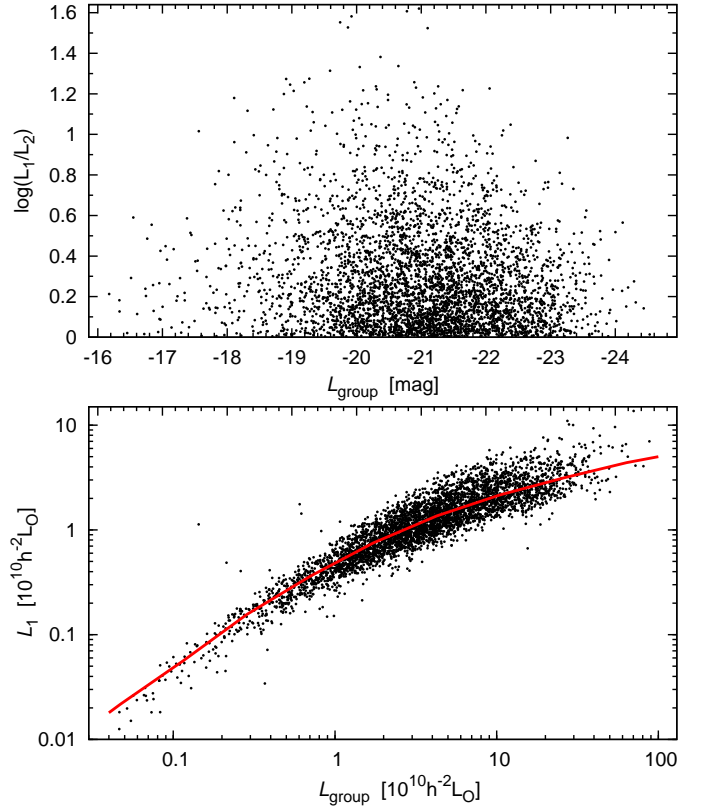}}
\caption{Upper panel: The luminosity gap between the first-ranked and
  second-ranked galaxies in groups, $\log L_1-\log L_2$, as a function
  of the first-ranked galaxy luminosity.  The lower panel shows the
  luminosity of the first-ranked galaxy as a function of the total
  estimated luminosity of the group.  The red line shows the median
  luminosity.}
\label{fig:19}
\end{figure}

\citet{hansen:07,yang:08a} found that luminosities of
  first-ranked galaxies of rich groups have a relatively small
  dispersion (see Fig.~5 of \citet{hansen:07} and Fig.~2 of
  \citet{yang:08a}).  In their halo occupation model
  \citet{cooray:05a} use for central galaxies a double-power-law model
  with a sharp decline at low luminosities, depending on the mass of the
  halo. Thus new data confirm the earlier results by
\citet{hubble:31, hubble:36, sandage:76, postman:95} and many others.
In most of these studies only very rich groups were considered.
  In our study also poorer groups were investigated, and we found that
  they have a lower low-luminosity limit in dense environments than rich
  groups do.  \citet{cooray:05a, yang:08a} showed that the median
luminosity of first-ranked galaxies depends strongly on the mass of
the halo (group).  To compare with their results we plot in the lower
panel of Fig.~\ref{fig:19} the luminosities of first-ranked galaxies
as a function of the estimated group total luminosity.  The median
luminosity of first-ranked galaxies is shown by a red line.  Our
results are very close to those by Yang et~al., see their Fig.~6 (left
panel).  Yang et~al. use as argument the estimated group (halo) mass,
which is closely related to the estimated total luminosity.  Our study
shows also that the median luminosity and the width of the luminosity
distribution of first-ranked galaxies depend on the density of the
environment.

These results mean that first-ranked galaxies of groups located in a
dense environment have a rather well fixed {\em lower} luminosity
limit.  The decrease of the LF at low luminosities was noticed in the
NGC901/902 supercluster by \citet{wolf:05} for dust-free old galaxies
(their Fig.~11). 

One of important results of the present study is the finding about the
nature of isolated galaxies in a flux-limited sample: most isolated
galaxies are actually the first-ranked galaxies, where the fainter
members of groups lie outside the visibility window.  A similar result
was obtained by \citet{yang:08a} using the halo occupation
model. Arguments used in our study and by Yang et~al. are very
different, so both studies complement each other.

\citet{yang:08a} also studied the gap between the first-ranked and
second-ranked galaxies. Their results show that the width for the gap
lies in range $\log(L_1/L_2)=0.0$--$0.6$. For our groups, the width of
the gap is even larger (see the upper panel of Fig.~\ref{fig:19}).
This Figure shows that the gap has the highest values for medium rich
groups of a total expected luminosity about $L_\mathrm{group} =
2\times 10^{10}~h^{-2} \mathrm{L}_\odot$, i.e. for groups of the type
of the Local Group. \citet{milosavljevic:06} used the luminosity
  gap statistic to investigate the cluster merger rate and to define
  ``fossil'' groups.

\citet{cross:01} determined a bivariate brightness distribution of a
subsample of 2dFGRS, i.e. the joint surface brightness-luminosity
distribution. Their analysis shows that if the surface brightness is
taken into account, then more exact extrapolation of the expected
total galaxy luminosity is possible. This results in a shift of the
characteristic magnitude $M^*$ brightwards by 0.33\,mag, and in an
increase of the total estimated number density of galaxies by a factor
of about 1.2.  The normalization of the LF for the Millennium Galaxy
Catalogue was discussed by \citet{liske:03,cross:04,driver:05}, where
normalization parameters for various previous determinations of the LF
were found.  In this paper we are comparing the LFs of first-ranked
galaxies, second-rank, and isolated galaxies. As the shifts found by
Cross et~al. and Liske et~al. influence these populations
approximately in the same manner, we expect that our results are
insensitive to the surface brightness effect.

An extensive study of LFs in various environments using the 2dF Galaxy
Redshift Survey was made by \citet{croton:05}.  Their main results are
the same that we found. Using the halo occupation model
  \citet{mo:04} and \citet{tinker:08} argue that the dependence of the
  LF on the large-scale 
  environment is determined by differences in the masses of DM haloes.  Our
  simple model shows that in different global environments the masses of
  DM haloes may differ by orders of magnitude, see Fig.~\ref{fig:18}, in
agreement
  with Mo et al. and Tinker \& Conroy results.
Semi-analytic models also predict that void
galaxies should be fainter than galaxies in dense regions \citep[][see
also \citet{einasto:05b}]{benson:03, tinker:08}.

In general, all studies show that as we move from high density
regions to low density regions (voids), galaxies become
fainter.  Interestingly, our study shows that in high density
environments (supercluster core regions), all LFs are
different from those in other environments. This result
is new, 
although the decrease of the LF at low luminosities was noticed in the
NGC901/902 supercluster by \citet{wolf:05} for dust-free old galaxies.
It is difficult to explain this effect with selection effects
since, as seen also in Fig.~\ref{fig:16}, the galaxies in supercluster
core regions in the area plotted here are located at the same distances
as galaxies in the nearby lower density regions where also
low luminosity galaxies are seen.
We believe that in cores of rich superclusters the faintest galaxies have been
swallowed by the brightest galaxies/groups, since in high density environments
merger events are much more common.
The reason why this has not been found in other studies,
is probably a different definition of the high density environment. Cores of
rich
superclusters are specific regions 
that contain clusters and groups of galaxies, as well as isolated
galaxies, and may contain X-ray clusters of galaxies.
 The morphology of supercluster cores differ from the morphology
 of supercluster outskirts
\citep{einasto:08}. Thus supercluster cores are
not just an environment that
contains groups of galaxies (groups may be located also in poor
superclusters), 
and the LFs of galaxies in supercluster cores
are not the same as the LFs of galaxy clusters
(see, for example, \citet{hansen:07}).  
The supercluster environment (excluding cores) can also
be considered as a high density environment
but here the galaxy and group content differs from that
in core regions. Thus here the definition
of the environment is crucial.

Our study does not include very faint galaxies, as, for
example, the study of the core region of the Shapley supercluster
\citep{mercurio:06}. Thus future work is needed to understand the
difference between the faint ends of the LFs in our and Mercurio et~al.
study.

One difference between the earlier studies and our work lies in the use of the
analytical LFs. Most earlier authors have used the Schechter
function, but our results show that this function is not good to describe the
bright end of the LF.  This difficulty was noticed already by
\citet{blanton:05} using the SDSS data. They also showed that the Schechter
function does not fit the LF of extremely low luminosity galaxies. In a
Shapley Supercluster region \citet{mercurio:06} conclude that the bright end
of the Schechter function is not sufficient to fit the data.  \citet{yang:08a}
also use different analytical expressions of LFs for different populations: a
log-normal distribution for first-ranked galaxies, and a modified
Schechter function for satellite galaxies.  The difficulty of the use of the
standard Schechter function for satellite galaxies lies in the fact that the
slope of the LF at high luminosities is much steeper than in the standard case
where the slope is fixed by the exponential law.  The double-power-law LF
overcomes both difficulties and can be used for brightest as well as
for satellite galaxies.  This difference is crucial in cases where only very
bright galaxies are visible, at the far end of flux-limited samples.  Here
small differences in the accepted analytical LF can lead to large differences
in the expected total luminosities of groups.

\citet{hoyle:05} studied the SDSS void galaxies. Their faint end slope
of the LF is comparable to our results ($\alpha=-1.0$--$-1.3$). They
also conclude that the faint-end slope is not strongly dependent on
the environment, at least up to group densities. This is in agreement
with our results, where the faint-end slope is almost the same for all
populations, except for the first-ranked galaxy.  However, in our
samples there are still small changes when moving from voids to
superclusters: the faint-end slope is steeper for void galaxies, and
becomes flatter when moving toward higher densities.  Our faint-end
slope $\alpha$ for the galaxy LF is in range $-1.0$--$-1.3$ (except
for the first-ranked galaxy), that is in good agreement with
observations and models \citep[see
e.g.][]{baldry:05,xia:06,khochfar:07,liu:08}.

\subsection{Interpretation of the luminosity function}

By definition, the transition of the power laws from low luminosities
to high luminosities occurs at a luminosity approximately equal to the
characteristic luminosity $L^*$, see Fig.~\ref{fig:13} and
Table~\ref{Tab4}.  The luminosity $L^*$ corresponds also to the median
luminosity of first-ranked galaxies when averaged over various
environments (except the void environment), see Fig.~\ref{fig:7}.
Fig.~6 and Table~1 of \citet{yang:08a} show similar coincidences for
groups (haloes) of various masses.

\citet{cooray:05c} demonstrated that the LF of galaxies can be
  calculated in the halo model using two premises: 1) the luminosities of
central
  galaxies have a lognormal distribution, $L^*$ being the mean
  luminosity of central galaxies in massive haloes; and 2) the
  luminosities of satellite galaxies are distributed as a power law.
  These assumptions mean that the high-luminosity section of the LF is
  determined by the first-ranked galaxies, and the low-luminosity
  section -- by the conditional LF of luminosity differences of satellite
galaxies 
  from the luminosity of the  brightest galaxy.

Properties of the LF of various types of galaxies in different
environments can be interpreted by differences of galaxy and group
evolution. In supercluster cores rich groups form through many
mergers, thus the second-ranked galaxies have been brightest galaxies
of poorer groups before they have been absorbed into a larger
group. In lower-density environment the merger rate is lower and
groups of galaxies have been collected only from nearby regions
through minor mergers and continuous infall of matter to
  galaxies as suggested by \citet{white:78}. Gas infall to galaxies
  (haloes) is very different in various environments, as shown by
  hydrodynamical simulations by \citet{keres:05}.  The satellite LF contains
  also information on the galaxy formation feedback, see \citet{cooray:05b}.

\section{Conclusions}\label{sec:7}

We used the 2dF Galaxy Redshift Survey to derive the LF of different
samples: the brightest (first-ranked ), second-ranked, satellite and
isolated galaxies and the LF of groups. We studied the LFs for various
environments.  The principal results of our study are the following:

\begin{itemize}

\item{} The LFs of galaxies (for all samples) are strongly dependent on the
  environment, in agreement with earlier studies.

\item{}  In the highest density regions (supercluster cores) the LFs
  for all galaxy populations have a well-defined lower luminosity
  limit, about $10^{9}\mathrm{L}_{\sun}$.  Here the first-ranked
  galaxies have larger luminosities than the first-ranked galaxies in
  other regions, in concordance with several earlier studies.

\item{} In the lowest density regions (voids) the LFs are shifted in
  respect to the LFs of all other regions, toward lower
  luminosities. Here, and in filament regions, the LFs of first-ranked
  galaxies have a plateau at the faint end.

\item{} The LF of second-ranked galaxies in high-density regions is
  similar to the LF of first-ranked galaxies in lower-density regions.
  The bright end of the LF of satellite galaxies is almost identical
  with the bright end of the LF of second-ranked galaxies. At lower
  luminosities the LF of satellite galaxies lies higher than the LF of
  second-ranked galaxies.

\item{} Almost all bright isolated galaxies can be identified with
  first-ranked galaxies where the remaining galaxies lie outside the
  observational window used in the selection of galaxies for the
  survey. Truly isolated galaxies are rare; they are faint and are
  located mainly in voids.

\item{} The LF of galaxies and groups can be expressed by a
  double-power-law more accurately than by the Schechter function.
  The biggest differences are in the bright end of the LF, where the
  Schechter function predicts too few bright galaxies. The advantage
  of double-power-law is clearly visible for the LF of groups.

\end{itemize}

\begin{acknowledgements}
  We are pleased to thank the 2dF GRS Team for the publicly available final
  data release. We thank the referee for stimulating
    suggestions.  The present study was supported by Estonian Science 
  Foundation grants No. 6104, 6106, 7146 and Estonian Ministry
  for Education and Science by grant SF0060067s08. This work has also
  been supported by the University of Valencia through a visiting
  professorship for E.~Saar and by the Spanish MCyT project
  AYA2003-08739-C02-01.  J.~Einasto  thanks Astrophysikalisches Institut Potsdam
  (using DFG-grant 436 EST 17/2/06), and the Aspen Center for Physics for
  hospitality, where part of this study was performed.
\end{acknowledgements}


\end{document}